# Simple Learning Rules Generate Complex Canonical Circuits

Joseph Olson[1] and Gabriel Kreiman[2,*]

[1]Department of Physics, Harvard University
[2]Children's Hospital, Harvard Medical School

[*]To whom correspondence should be addressed at:
gabriel.kreiman@tch.harvard.edu

Abstract

Cortical circuits are characterized by exquisitely complex connectivity patterns that emerge during development from undifferentiated networks. The development of these circuits is governed by a combination of precise molecular cues that dictate neuronal identity and location along with activity dependent mechanisms that help establish, refine, and maintain neuronal connectivity. Here we ask whether simple plasticity mechanisms can lead to assembling a cortical microcircuit with canonical inter-laminar connectivity, starting from a network with all-to-all connectivity. The target canonical microcircuit is based on the pattern of connections between cortical layers typically found in multiple cortical areas in rodents, cats and monkeys. We use a computational model as a proof-of-principle to demonstrate that classical and reverse spike-timing dependent plasticity rules lead to a formation of networks that resemble canonical microcircuits. The model converges to biologically reasonable solutions provided that there is a balance between potentiation and depression and enhanced inputs to layer 4, only for a small combination of plasticity rules. The model makes specific testable predictions about the learning computations operant across cortical layers and their dynamic deployment during development.

## Introduction

Neocortical circuits constitute the fundamental building blocks for cognitive computations and are characterized by a bewildering complexity in connectivity patterns. How such intricate and precise connectivity arises through development and learning constitutes a fundamental challenge for neuroscience. In part, the answer relies on a web of molecular cues that guide neuronal precursors to specific brain areas (e.g., specifying which neurons will end up in primary visual cortex versus olfactory cortex), and to specific layers within those areas (e.g., specifying which neurons will reside in layer 4 versus layers 2/3) (Bolz et al. 1996; Castellani and Bolz 1997; Callaway 1998b; Larsen and Callaway 2006; Lui et al. 2011; Silbereis et al. 2016). In addition to molecular cues, activity- dependent mechanisms play a central role in shaping and/or refining neural circuits, both during development and subsequent learning (Feldman and Brecht 2005; Fox and Wong 2005; Karmarkar and Dan 2006; Butts et al. 2007; Espinosa and Stryker 2012; Bennett and Bair 2015; Lim et al. 2015).

Here we investigate how simple activity-dependent mechanisms can give rise to complex circuit structures by adequately modifying the strength of neuronal connections. An important activity-dependent mechanism governing the connection strength between neurons is spike-timing dependent plasticity (STDP) (Markram et al. 1997; Bi and Poo 1998). Different forms of STDP have been observed throughout biological circuits (for reviews, see Abbott and Nelson 2000; Caporale and Dan 2008; Froemke et al. 2010). We consider two specific forms of STDP that have been widely observed in cortex: classical STDP (cSTDP, Fig. 1a-b top) and reverse STDP (rSTDP, Fig. 1a-b bottom). In cSTDP, long-term potentiation (LTP) strengthens connections when a pre-synaptic action potential precedes a post-synaptic action potential while long-term depression (LTD) weakens connections when the post-synaptic action potential precedes the pre-synaptic action potential (Markram et al. 1997; Bi and Poo 1998; Debanne et al. 1998; Feldman 2000; Sjöström et al. 2001; Froemke et al. 2005). cSTDP can be thought of as a mechanism that promotes causally linked feedforward connections. In rSTDP, connection strengths change in the opposite direction: LTD weakens connections when a pre-synaptic action potential precedes a post-synaptic action potential while LTP strengthens connections when the post-synaptic action potential precedes the pre-synaptic action potential (Letzkus et al. 2006; Sjöström and Häusser 2006; Burbank and Kreiman 2012). rSTDP can be thought of as a mechanism that promotes feedback connections. Fig. 1b schematically illustrates connection becoming stronger or weaker depending on the relative timing of the pre/post-synaptic spikes and the STDP rule. The assignment of learning rules across connections can have a major impact on the resulting structure of a neural circuit. For instance, computational simulations show that cSTDP leads to the elimination of loops in fully connected networks (Kozloski and Cecchi 2010) and rSTDP enhances feedback connections in a multiple-layer network (Burbank and Kreiman 2012). We extend these ideas by investigating whether it is possible to generate complex connectivity patterns such as those observed in neocortical circuits purely from activity-dependent mechanisms based on STDP and starting from all-to-all connectivity.

We focus on the approximately canonical inter-laminar connectivity observed in neocortical circuits. Such connectivity has been observed in macaque V1 (Callaway 1998a, Fig. 2) and other visual cortical areas (Felleman and Van Essen 1991), in cat V1 (Douglas and Martin 2004, Fig. 1) and in mice (Larsen and Callaway 2006). The target canonical circuitry of inter-laminar connections is simplified to the structure in Fig. 1c. This simplified circuitry ignores significant aspects of neocortical circuits including sub-laminar structure such as horizontal connections within a layer (Binzegger et al. 2004), sub-divisions of layer 4, distinctions between layers 5 and 6, different neuronal types within each layer, and real-valued connection strengths that are not 0 or 1 (see Discussion). To a reasonable first-order simplification, the inter-laminar connectivity pattern is conserved across multiple cortical regions and even across species. We start with a spiking network that contains 3 layers, labeled layer 4, layer 2/3 and layer 5/6. These layers are initially connected all-to-all and connections undergo either cSTDP or rSTDP (Fig. 1d). We investigate which combinations of STDP-based learning rules give rise to connections that match the target circuitry. We demonstrate that it is possible to rapidly develop a good approximation to the target canonical circuit in Fig. 1c from the initial random circuit in Fig. 1d using a small cluster of configurations of simple activity-dependent STDP learning rules.

## Results

We asked whether it is possible to develop complex architectures with connectivity similar to that of neocortical circuits starting from fully connected neurons distributed into three layers and following simple STDP rules: classical and reverse STDP. We consider as a target the idealized version of a canonical microcircuit schematically illustrated in Fig. 1c. This circuit is an abstraction of the inter-laminar connectivity in cortical areas reported in macaque, cats, and mice (Callaway 1998a, Douglas and Martin 2004, Larsen and Callaway 2006). In the simplified version of biological connectivity considered here, connections are either maximally strong (strength of 1) or absent (strength of 0) and only the main connections are represented (see Discussion). In the initial conditions for the developmental simulations, all neurons in one layer are connected to all neurons in another layer and all weights are initialized to 0.5. Each layer contains 33 integrate-and-fire neurons (see Supplementary Table S1 for simulation parameters). In each simulation and for each pair of layers and connectivity direction (e.g. neurons from layer 4 projecting to layer 2/3), we consider a specific learning rule (cSTDP or rSTDP) governing how the weights evolve for all the corresponding synapses. Because there are 9 different types of connections (3 types of within-layer connections plus 6 types of between-layer connections), there is a total of $2^9 = 512$ different configurations (we refer to a configuration as a particular combination of cSTDP or rSTDP for each connection type). Fig. 1d (expanded in Supplementary Fig. S1) shows one of those possible configurations. Each neuron receives excitatory input from independent homogenous Poisson neurons ($E_4$, $E_{2/3}$, and $E_{5/6}$). Layer 4 is assumed to receive more excitatory input than layers 2/3 and 5/6 (i.e. $E_4 > E_{2/3}, E_{5/6}$) because it is typically the layer receiving input from the thalamus or from earlier cortical areas (Felleman and Van Essen 1991). Additionally, each neuron receives inhibitory input from independent Poisson neurons whose firing rates change as a function of the fraction of active integrate-and-fire neurons. The STDP curves are modeled as two exponential functions with amplitudes $A_+$ and $A_-$, and time constants $\tau_+$ and $\tau_-$ (see Methods for details and Supplementary Table S1 for parameter values). Each configuration was simulated n=5 times for 60 seconds. After stable equilibrium was reached, usually well before 60 seconds, weight fluctuations remained small compared to the weight values (Supplementary Fig. S2). At the end of each simulation, we averaged the weights into a 3 × 3 weight matrix $W$.

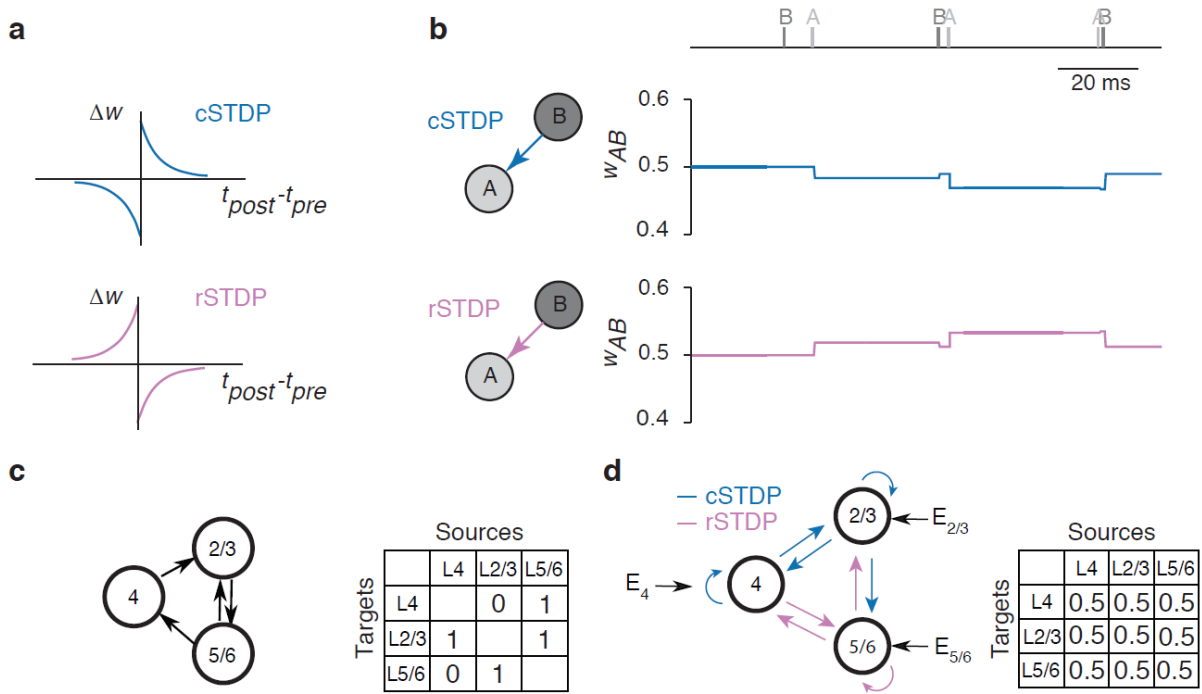

Figure 1
**Model description. a,** Schematic illustration of how the change in synaptic weights depends on the relative timing of pre- and post-synaptic spikes for classical STDP (top) and reverse STDP (bottom). **b,** Sample spike trains from two neurons, A and B (top), and how the synaptic weight from B to A ($w_{AB}$) evolves with the occurrence of each spike under cSTDP (middle) or rSTDP (bottom). **c,** Schematic of target connectivity in the canonical circuit, simplifying the inter-laminar connectivity patterns found in cortical circuits in rodents, cats and monkeys. There are 3 layers (L4, L2/3 and L5/6); the direction of the arrows denotes the desired connectivity. The connections are idealized in the connectivity weight matrix shown on the right where row $i$, column $j$ is 1 iff there is a connection from column $j$ onto row $i$ (see Methods). **d,** Example initial conditions where all weights start at 0.5. Each layer receives external excitatory inputs ($E_4, E_{2/3}, E_{5/6}$) in addition to recurrent inputs within the same layer and inputs from other layers. A specific plasticity rule was assigned to each of the 9 possible connections between or within layers (see Methods). The combination of learning rules depicted here is only one of the 512 possible combinations examined throughout this study.

### Some configurations develop into networks that resemble the target microcircuit

Fig. 2 shows one STDP configuration that leads to a network resembling the target microcircuit and one that does not. The final weights for the circuit in Fig. 2a approximate the target matrix for the idealized network in Fig. 1c. We compared the final weight matrix $W$ with the target matrix $T$ by defining the degree of success of each configuration as $s = 1 - 6^{-\frac{1}{2}}\|W - T\|_F$ where $\|\cdot\|_F$ is the Frobenius matrix norm. The diagonal elements, corresponding to the within-layer weights, do not contribute to the success metric (see Discussion). Since weights are bounded between 0 and 1, $s$ is bounded between 0 and 1 with $s = 1$ if and only if $W = T$. The configuration in Fig. 2a has a success of $s = 0.70 \pm 0.01$. In contrast, the configuration in Fig. 2b has a success of $s = 0.22 \pm 0.01$. The initial condition has a success $s = 0.5$, hence the configuration in Fig. 2a develops into a circuit that becomes more similar to the target whereas the configuration in Fig. 2a develops into a circuit that is even less similar to the target than the initial conditions.

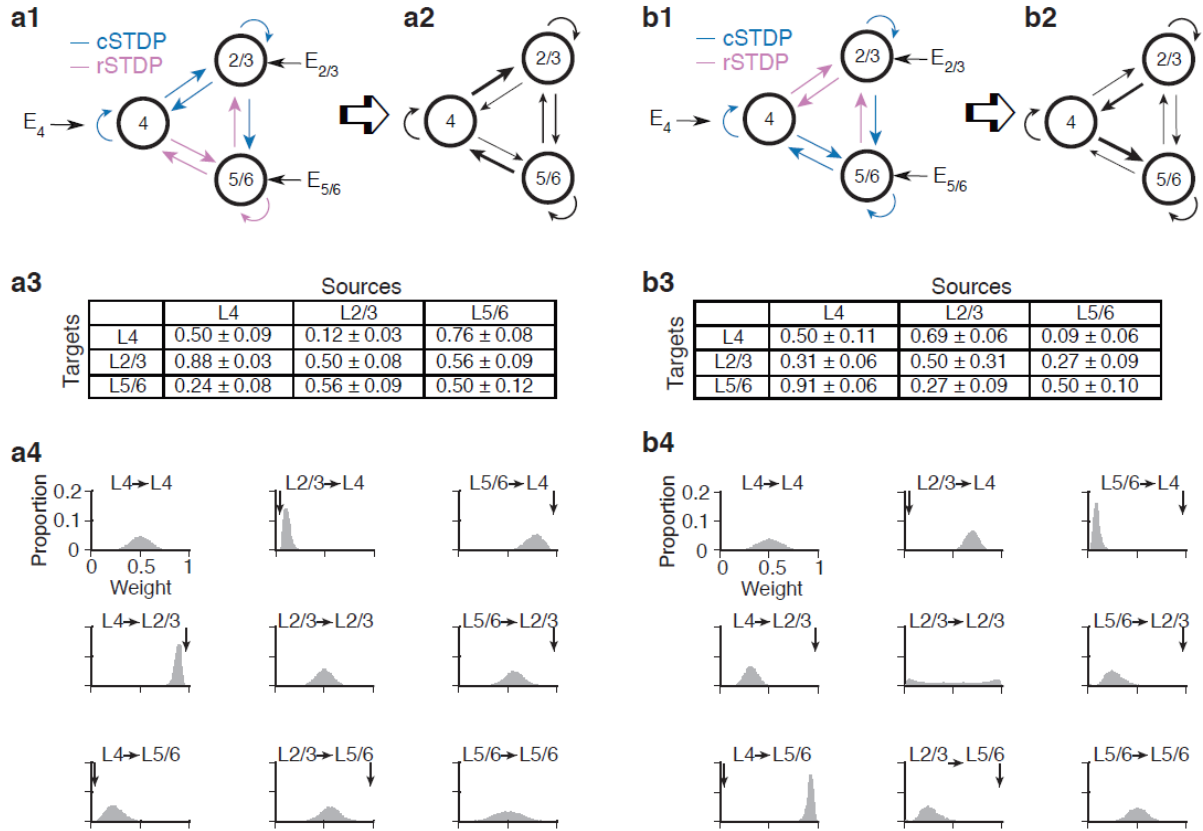

Figure 2
**Two example simulations, one successful (a), one not (b). a1/b1,** Initial configuration. **a2/b2,** Network at the end of the simulation. Line widths are proportional to the corresponding weights. **a3/b3,** Weight matrices at the end of simulation, repeated 5 times (mean $\pm$ SD across neurons, $n = 33 \times 33 \times 5 = 5{,}445$, averaged over the last 5 seconds of simulations, see Methods). **a4/b4,** Histograms showing the distribution of weights for each pair of layers.

### The best configurations share a specific combination of learning rules

We computed the degree of success for each of the 512 possible learning rule configurations (Supplementary Fig. S3). The degree of success ranges from $s = 0.14 \pm 0.01$ (worst) to $s = 0.70 \pm 0.01$ (best) (Fig. 3d). The weights and success of the best 16, middle 16, and worst 16 configurations are shown in Supplementary Table S2. For most configurations, the degree of success is lower than that of the initial conditions (Fig. 3d), i.e., most combinations of learning rules do not lead to the formation of circuits resembling the target one. Interestingly, in order for the model to arrive at an architecture that resembles the target canonical circuit, the plasticity rules between layers need to be within a certain configuration of cSTDP/rSTDP rules (Fig. 3a). Other combinations of cSTDP and rSTDP led to different architectures (e.g. Fig. 2b, 3d, 4a, Supplementary Fig. S3). Specifically, the model predicts that connections $L4 \rightarrow L2/3$ and $L2/3 \rightarrow L4$ both follow cSTDP; connections $L4 \rightarrow L5/6$ and $L5/6 \rightarrow L4$ both follow rSTDP; and connections $L5/6 \rightarrow L2/3$ follow rSTDP. The connection $L2/3 \rightarrow L5/6$ formed equally well with either cSTDP and rSTDP (Fig. 4b). Altogether there are 4 unspecified connections among the best $2^4 = 16$ configurations. A configuration is in the best 16 if and only if it shares the combination of rules specified above and illustrated in Fig. 3a. Furthermore, the best 16 configurations are separated from the rest by a gap in the success curve (Fig. 3d, Supplementary Fig. S3a). A similar gap separates the worst 8 which also display a common configuration of STDP learning rules (Supplementary Fig. S3a, Supplementary Table S2). The combinations of learning rules shown in Fig. 3a for the best 16 configurations, lead to the average weights shown in Fig. 3c and the circuit depicted in Fig. 3b, which resembles the target canonical circuit in Fig. 1c.

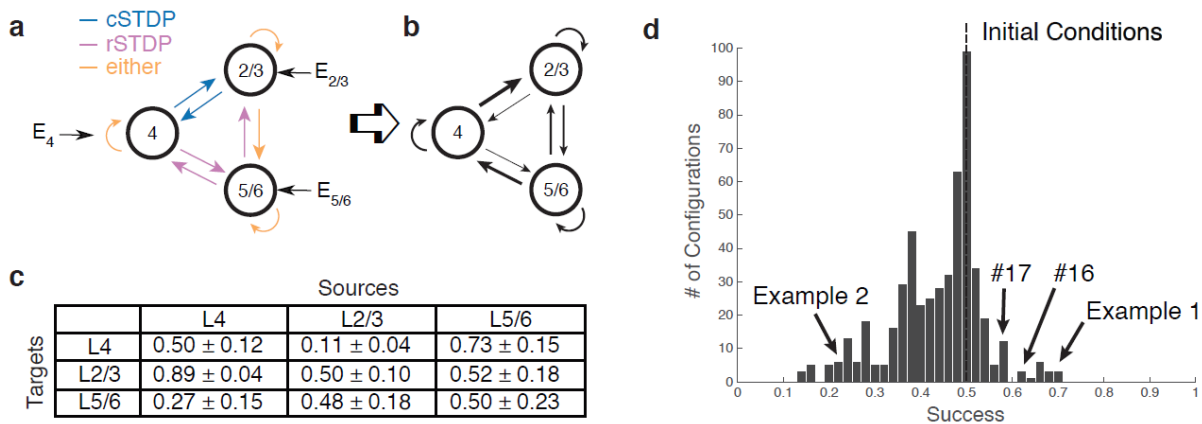

Figure 3
**Configuration for the best 16 models. a,** Learning rules for each connection for the best 16 models. **b,** Final circuit at the end of the simulations, averaged across the best 16 models. **c,** Final weights for the best 16 models ($n = 5,445 \times 16 = 87,120$). **d,** Average success of each of 512 configurations ($n = 5$). Example 1 is the configuration shown in Fig. 2a and Example 2 is the configuration shown in Fig. 2b. Also shown is the success, 0.5, of the initial conditions. Note the gap between configuration number 16 and configuration number 17, as well as the gap before the worst 8 simulations.

**Models with only one type of learning rule between layers outperform models with mixed learning rules**

The previous results assume that all the connections from one layer to another follow the same learning rule. In order to evaluate the impact of this assumption on the results, we systematically consider each pair of layers and vary the fraction of connections following cSTDP from none to all (Fig. 4a-b, Supplementary Fig. S4). For example, in Fig. 4a, we vary the fraction of cSTDP connections $L5/6 \rightarrow L2/3$, such that 0% cSTDP (100% rSTDP) corresponds to one of the 16 best configurations (arrow in Fig. 4a, right). The success value decreases monotonically as more cSTDP connections are added, departing from the best configuration. At 100% cSTDP, success drops to almost the initial condition value. In contrast, success is essentially unperturbed while varying the fraction of cSTDP connections $L2/3 \rightarrow L5/6$ (Fig. 4b), further confirming that either learning rule is adequate for the connections between these two layers.

We vary the fraction of cSTDP connections between each pair of layers in the best 16 configurations. In each case, success peaks when models have either 100% cSTDP or 100% rSTDP, matching one of the configurations in the best 16 group (Supplementary Fig. S4). The right column in Supplementary Fig. S4 shows large error bars because the configurations considered in these averages, having cSTDP connections $L2/3 \to L5/6$, come from both the higher and lower ends of the best 16 ranking (Supplementary Table S2).

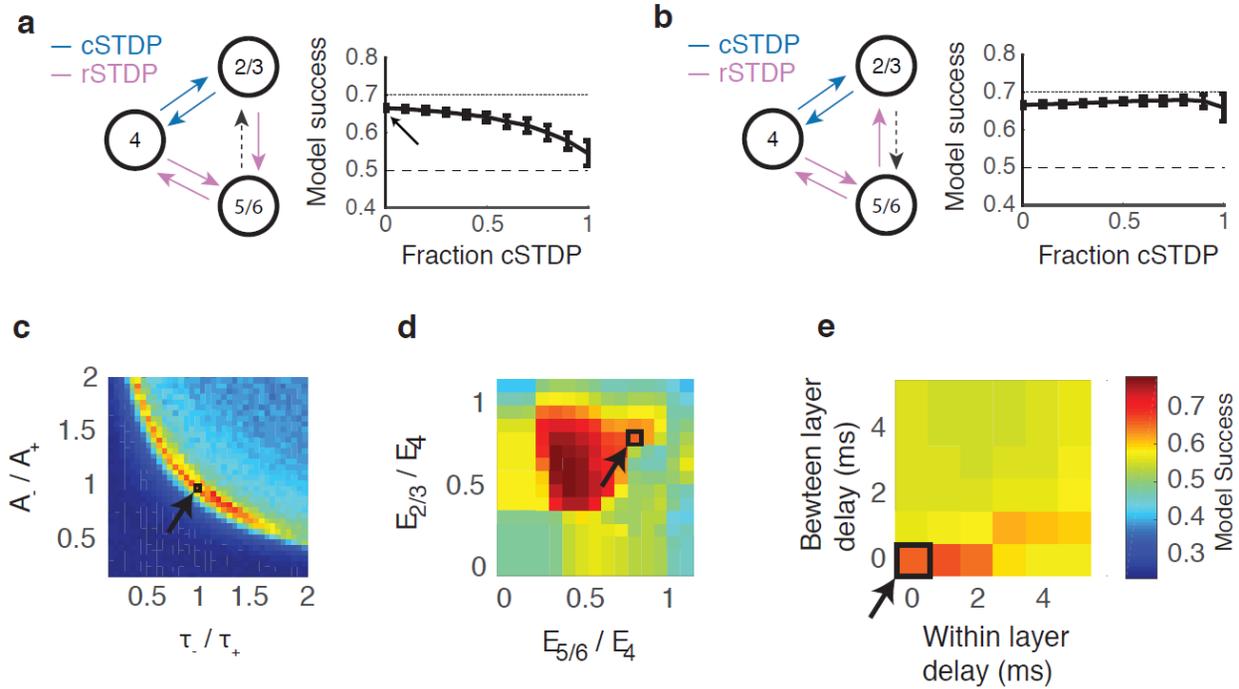

Figure 4

**Robustness of the best configurations. a-b,** We vary the fraction of cSTDP connections from 0 to 1 (all rSTDP to all cSTDP) from layer 5/6 to layer 2/3 (**a**) or from layer 2/3 to layer 5/6 (**a**). The connection shown as a dashed arrow is the one that is subject to different fractions of cSTDP. The model success curve is averaged across 5 simulations and across within-layer connections (8 possible configurations) for a total of $n = 40$. Error bars represent standard deviations. The dashed line shows the model success for the initial conditions. The dotted line shows the model success for the overall best configuration, which is depicted in Fig. 2a. The arrow in (**a**) points to the default condition corresponding to best 16 configurations. In (**b**), where both extremes correspond to best 16 configurations, cSTDP and rSTDP lead to equivalent model success. **c,** Model success (color scale shown on right) for different combinations of STDP amplitude and time constant ratios. The arrow points to the default condition. Success is averaged across 5 simulations and across the best 16 configurations for a total $n = 80$. **d,** Model success for different ratios of excitatory inputs ($n = 80$). **e,** Model success for different combinations of within and between layer delays ($n = 80$).

### The formation of the target microcircuit depends on the balance between potentiation and depression

Next, we examine the robustness of the conclusions to several of the critical parameters and assumptions in the simulations. In Fig. 4c, we vary the STDP exponential parameters $A-$ and $\tau-$ away from their default values $A_- = A_+$ and $\tau_- = \tau_+$. There is a sharp decrease in success away from the curve defined by $A_+\tau_+ = A_-\tau_-$. The quantities $A_+\tau_+$ and $A_-\tau_-$ correspond to the area under the positive and negative parts of the cSTDP curve in the best part of Fig. 1a, and conversely, the area under the negative and positive parts of the rSTDP curve. The decrease in success is due to weights strengthening and weakening as a result of a bias towards potentiation or depression. Setting $A_+\tau_+ > A_-\tau_-$ leads to enhanced strengthening/weakening of connections following the cSTDP/rSTDP rules respectively. Conversely, setting $A_+\tau_+ < A_-\tau_-$ leads to enhanced weakening/strengthening of connections following the cSTDP/rSTDP rules respectively. As an example of a failure mode, increasing $A_-\tau_-$ results in strong connections that follow rSTDP from layer 4 to layer 5/6 whereas the target circuit has none of those connections.

**The formation of the target microcircuit depends on increased inputs to layer 4**

In the models described so far, the external inputs to layer 4 ($E_4$) are stronger than the external inputs to the other two layers ($E_{2/3} = E_{5/6} = 275, E_4 = 350$). We examined the impact of the relative external input strengths on the degree of success of a model by varying $E_{2/3}$ and $E_{5/6}$ (Fig. 4d). Consistent with the assumption that layer 4 is the main input layer, there is a sharp decrease in success for models with $E_{2/3} > E_4$ or $E_{5/6} > E_4$. In contrast, as the amount of input to layer 4 increases in comparison to layers 2/3 and 5/6, the degree of success also increases. Supplementary Fig. S3 depicts the degree of success for all 512 configurations under two such conditions with different levels of $E_4$ inputs. Some configurations in these models with smaller $E_{2/3}/E_4$ and $E_{5/6}/E_4$ ratios show large degrees of success close to 1 (e.g. best configurations in Supplementary Fig. S3. Additionally, these models with enhanced $E_4$ inputs also show increased separation for the best models from the rest (Supplementary Fig. S3). However, as $E_4$ increases, there is also a decrease in the average equilibrium firing rates in layers 2/3 and layers 5/6 (Supplementary Fig. S3).

Conversely, when $E_{2/3}$ and $E_{5/6}$ are enhanced, there is a decrease in success. This is because as $E_{2/3}$ and $E_{5/6}$ get close (or even surpass) $E_4$, there is no longer a driving force into layer 4. We investigated further the case where after circuit development, the enhanced driving force into layer 4 is taken away and all inputs are equal (Supplementary Fig. S5). In this case, the structure of the circuit vanishes and the circuit adapts to reflect the symmetry in the inputs with the weights converging towards 0.5.

Note that the strength of the external inputs into a layer depends the number of connections as well the weights which undergo cSTDP. However, the variability of the external excitatory neuron's spike statistics leads to the same weight values from the external populations into each layer. The average final weights into layer 4, 5/6, and 2/3 from their respective external inputs are 0.53±13, 0.52±15, and 0.53±14. Thus, the number of external input connections determines the strength of the input.

**Long delays between layers disrupt the development of the target microcircuit**

In the simulations reported so far, synaptic transmission was considered to be instantaneous, i.e., a spike in one neuron exerted an immediate effect on its post-synaptic target. We evaluate the consequences of introducing delays between layers (Fig. 4e). The degree of success remains high for short synaptic delays of up to 2 ms between neurons within the same layer. Outside of this regime, introduction of delays disrupted the success of the simulations.

**Early development of L5/6 to L4 connections disrupts the development of the target microcircuit**

In the simulations presented thus far, the architecture and STDP rules were established from the onset and all the connections started to change at the same time. The ensuing dynamics for the different inter-laminar connections were similar, and they achieved their final values approximately at the same time (Fig. S2b). We next considered scenarios in which one of the six inter-laminar connections arose before the others to evaluate whether the development of the target circuit was influenced by the order in which connections solidified. We ran the simulations while fixing each of the 6 inter-laminar connections separately to the final weight value obtained in the default simulations (Fig. 3c) while all the other connections changed according to the corresponding STDP rules. When the weights from L5/6 to L4 were fixed to 0.73 from the beginning, the network was unable to converge the target circuit (Supplementary Fig. S6). However, in all other cases when one of the connections was pre-determined, the network was able to converge to the target circuit (Supplementary Fig. S6).

## Discussion

We asked whether simple plasticity rules can give rise to the rich connectivity patterns of canonical circuits in neocortex. Starting from a fully connected 3-layered network, we demonstrate that a simple combination of spike-timing dependent plasticity (STDP) rules can rapidly lead to a complex architecture which captures some of the essential connectivity patterns of cortical circuits. The proposed model follows the essential ingredients of previous work with spiking networks undergoing plasticity including integrate-and-fire neurons, STDP, 'tabula rasa' initial conditions, and biologically plausible parameters (Abbott and Nelson 2000; Kozloski and Cecchi 2010; Burbank and Kreiman 2012). The model leads to a stable (Supplementary Fig. S2) and robust solution (Fig. 3) that resembles a simplified version of the canonical circuit (Fig. 1c), provided that the connections respect a specific combination of cSTDP and rSTDP rules (Fig. 3, Supplementary Fig. S3), provided that there is a balance between potentiation and depression ($A_+\tau_+ \approx A_-\tau_-$, Fig. 4c), and provided that there are stronger external inputs to layer 4 (Fig. 4d).

We compared the resemblance of the final states of our model to the target canonical circuit with a success metric. The success of our simulations does not reach 1.0, but this is to be expected for several reasons. First, the target canonical circuit is idealized to have connection strengths of 0 or 1 whereas real connections follow a distribution of synaptic strengths. Second, noise is continuously introduced into the circuit from the external Poisson spiking neurons so that the weights cannot reach a stable value of 0.0 or 1.0. Furthermore, the soft bounds imposed on the weights (see Methods) push the weight values away from 0.0 and 1.0, making it highly unlikely that weights would settle on those values. Third, although it is possible to fine tune parameters such that the models have a higher degree of success, e.g. Fig. 4e, our aim is not to reach success = 1, but rather to show as a proof-of-principle, that activity dependent mechanisms can build circuits qualitatively similar to those found in biological systems.

The success metric did not include the within-layer connections, because the relative strength of within layer connections compared to the between layer connections remains unclear. The within-layer connections do not contribute to the success metric because the average within-layer weight is consistently 0.5 during the entire simulation. This is because the within-layer weights all undergo the same type of STDP, they are initialized at 0.5, and potentiation of weight $w_{ij}$ is exactly the opposite of depression of $w_{ji}$. Although within-layer connections did not directly contribute to the degree of success of a configuration, they indirectly affected the weights of the between-layer connections. In the most concrete example, when STDP rules are configured as in the best 16 configurations with the additional constraints that both $L2/3 \rightarrow L5/6$ and $L5/6 \rightarrow L5/6$ follow cSTDP, multimodal weight distributions were observed. The weight distributions, averaged across these 4 configurations, are compared to those averaged across the other (unimodal) 12 configurations in Supplementary Fig. S7.

In order for the model to arrive at an architecture that resembles the target canonical circuit, the plasticity rules between layers need to be within a certain configuration of cSTDP/rSTDP rules (Fig. 3a). Interestingly, this configuration is consistent with experimental studies. Plasticity governed by cSTDP at proximal synapses and rSTDP at distal synapses of $L2/3 \rightarrow L5/6$ pyramidal neurons has been observed in rat primary somatosensory cortex (S1) (Letzkus et al. 2006; Sjöström and Häusser 2006). Furthermore, our results are consistent with a study which reports cSTDP from $L4 \rightarrow L2/3$ in rat S1 (Feldman 2000). Although our simulations do not make any strong predictions about STDP rules within layers, experimental studies have observed that connections within L2/3 and within L5/6 follow cSTDP (Markram et al. 1997; Egger et al. 1999), and connections within L4 follow rSTDP (Egger et al. 1999). Our model predicts that rSTDP may also be observed between connections $L5/6 \rightarrow L4$ and $L5/6 \rightarrow L2/3$.

The requirement for an approximate balance between potentiation and depression has also been proposed in previous studies of plasticity in spiking networks (Burbank and Kreiman 2012; Babadi and Abbott 2013). Consistent with these studies we see that unbalanced potentiation and depression can lead to unchecked strengthening or weakening of connections. While precise measurements of $A_+, A_-, \tau_+, \tau_-$ are difficult to come by, we estimated these quantities from different empirical STDP studies. Supplementary Fig. S8 shows that these estimates are approximately consistent with a balance between total potentiation and depression (the area under the curve above and below the y-axis in the STDP curves in Fig. 1a).

The second requirement is that the external inputs to layer 4 need to be stronger than those to other layers. This requirement is consistent with a large body of literature which indicates that cortical areas mostly receive input via layer 4. For primary sensory areas, this input comes from thalamus, and for higher sensory cortical areas this input comes from layer 2/3 of other cortical areas (Felleman and Van Essen 1991; Callaway 1998a; Miller 2003). It has recently been reported that layer 5/6 also receives direct input from the thalamus (Constantinople and Bruno 2013). As each layer in our model receives external input, this does not contradict our assumptions as long as the input to layer 4 is stronger. Our model is *not* specific to the thalamocortical system, though. As long as the external input to layer 4 is stronger, this model may also capture the formation of between layer connections in other cortical areas.

More is known about the development of primary cortical areas deriving inputs from the thalamus (e.g., primary visual cortex) than about other cortical areas (e.g. visual areas V2, V4, etc.). Early stages of primary cortical circuit development occur *before* thalamic afferents reach cortical layer 4. This observation has led many investigators to conclude that the development of between layer connectivity is primarily driven by molecular cues with the role of activity-dependent mechanisms confined to circuit refinement (Lund and Mustari 1977; Rakic 1977; Callaway 1998b; Pasko Rakic 2009). However, it is conceivable that the type of rapid restructuring of between layer connectivity proposed by this model might rely on inputs from a transient structure called the subplate, rather than on direct inputs from the thalamus. Positioned directly beneath developing cortical cells, the subplate is the target of early thalamic afferents where they wait for days (in rats) or weeks (in cats) before entering the cortical plate (Lund and Mustari 1977; Shatz and Luskin 1986). During this time, subplate neurons project to a developing layer 4 and are capable of firing action potentials (Allendoerfer and Shatz 1994) and are the first cortical neurons to respond to sensory stimuli (Wess et al. 2017). Taken together, it is possible that early spontaneous activity in the subplate, rather than thalamus, may drive developing cortical circuits by providing enhanced input to layer 4. Consistent with this notion, disruption of thalamocortical afferents results in largely intact laminar structure (Miyashita-Lin et al. 1999; Li et al. 2013), perhaps because in this preparation the thalamic projections to the subplate remained undisturbed.

The type of activity-dependent plasticity mechanism proposed here does not necessarily rely on actual sensory experience. For example, in the context of vision, the model does not require post-natal visual inputs and could well take place during the embryonic stage. The type of activity used in the current study contains no structure (beyond the enhanced inputs to layer 4). We speculate that richer and structured activity patterns, in combination with molecular cues, might lead to even more complex circuits. Indeed, the target canonical microcircuit considered here clearly constitutes a major oversimplification abstracting away much of the exquisite and enigmatic architecture of cortex, including the differentiation between six neocortical layers, the vast array of different types of excitatory and inhibitory neurons, the distance dependence in connectivity patterns, and the non-uniform distribution of synaptic inputs along dendrites, among many others. The current model clearly does *not* claim that every aspect of the cortical connectivity pattern can be purely generated by STDP. The model demonstrates that adequately combining very simple activity-dependent learning rules can rapidly lead to the emergence of complex circuits that capture essential principles of the cortical connectome.

# Methods
## Model description

All the models have the same overall structure, consisting of 99 integrate-and-fire neurons split evenly into 3 layers, 33 neurons per layer (Supplementary Fig. S1). We refer to those layers as 'layer 2/3' (L2/3), 'layer 4' (L4), and 'layer 5/6' (L5/6). The network is initially connected all-to-all (no self-connections) with weights set to 0.5, half the maximum value of $w_{max} = 1$. The weights are constrained to be non-negative and the bounds are imposed using a soft-max mechanism within the STDP update rule described in the section **Weight Changes**.

In addition to the input from the internal network described above, each neuron receives input from external excitatory Poisson neurons of firing rate 20 Hz. Each neuron in layer 4, layer 2/3, and layer 5/6 receive input from $E_4 = 350$, $E_{2/3} = 275$, $E_{5/6} = 275$ external excitatory Poisson neurons, respectively. Each layer has a separate pool of 2500 external excitatory neurons supplying input. Connections from the external population to each network neuron are drawn randomly. All neurons also receive external inhibition from 250 randomly selected neurons chosen from a pool of 1250 Poisson neurons. The inhibitory neurons had firing rates which track average network activity to provide excitatory/inhibitory balance for the network. The firing rate of these external inhibitory neurons, $r_{inh}(t)$, depends on the fraction of firing neurons in the network at time t, denoted by $\gamma(t)$. At each time step, dt = 0.1 ms, the rate is updated by $r_{inh}(t+1) = r_{inh}(t) + \gamma(t)(rmax - rmin)$ where $r_{inh}(0) = 20$ Hz, rmax = 1000 Hz, and rmin = 5 Hz. Also, $r_{inh}(t)$ decays exponentially every time step with a time constant of $\tau_I = 2$ ms, obeying $\tau_I \frac{dr_{inh}}{dt} = -r_{inh}$. See Supplementary Table S1 for a full list of parameters used in the simulations.

## Individual neuron dynamics

The simulations are based on networks proposed by (Song et al. 2000; Kozloski and Cecchi 2010). All simulations were run in MATLAB 2013b (Mathworks, Natick, MA) and all the code is available at http://klab.tch.harvard.edu. Each neuron's membrane potential is governed by

$$\tau_m \frac{dV_i}{dt} = V_{rest} - V_i + \sum_{j \in \{exc \to i\}} g_{exc}^{ij}(t)(E_{exc} - V_i) + \sum_{j \in \{inh \to i\}} g_{inh}^{ij}(t)(E_{inh} - V_i)$$

where j and i refer to pre-synaptic and post-synaptic neurons respectively, $\{exc \to i\}$ denotes the set of excitatory inputs to neuron i, $\{inh \to i\}$ denotes the set of inhibitory inputs to neuron i, $g_{exc}^{ij}(t)$ is the excitatory synaptic conductivity from j onto i at time t, $g_{inh}^{ij}(t)$ is the inhibitory synaptic conductivity from j onto i at time t, $\tau_m = 20$ ms, $V_{rest} = 60$ mV, $E_{exc} = 0$ mV, and $E_{inh} = 70$ mV. The set of excitatory inputs includes those from the external Poisson neurons as well as those from the internal network. The set of inhibitory inputs include only those from the external Poisson neurons. After the voltage reaches a threshold, $V_{thresh} = -54$ mV, the neuron spikes and the voltage is reset to $V_{reset} = -60$ mV.

## Weight changes

When a presynaptic spike occurs, the synaptic conductance is increased by an amount proportional to the synaptic weights: $g_{exc}^{ij}(t) = g_{exc}^{ij}(t-1) + \alpha w_{ij}(t)$ and $g_{inh}^{ij}(t) = g_{inh}^{ij}(t-1) + \alpha w_{inh}$ with $\alpha = 0.01$ and $w_{inh} = 1.5$. Otherwise, $g_{exc}^{ij}(t)$ and $g_{inh}^{ij}(t)$ decay exponentially with time constants $\tau_{exc} = \tau_{inh} = 5$ ms. All excitatory synaptic weights in the model are subject to plasticity (including those from the external excitatory inputs which are initialized at $w_{exc} = w_{max}$); all the inhibitory synaptic weights are fixed. Excitatory weights are updated by $w_{ij}(t) = w_{ij}(t-1) +$ $\Delta w_{ij}(t)$ where $\Delta w_{ij}(t)$ is determined by either classical STDP (cSTDP) or reverse STDP (rSTDP) rules. As depicted in Fig. 1a, the equations governing cSTDP and rSTDP are given by:

$$\text{cSTDP: } \Delta w_{ij}(t) = \begin{cases} A_+(1-w_{ij})^\mu e^{-\Delta t/\tau_+} & \text{if } \Delta t > 0 \\ -A_- w_{ij}^\mu e^{\Delta t/\tau_-} & \text{if } \Delta t < 0 \end{cases}$$

$$\text{rSTDP: } \Delta w_{ij}(t) = \begin{cases} -A_+ w_{ij}^\mu e^{-\Delta t/\tau_+} & \text{if } \Delta t > 0 \\ A_-(1-w_{ij})^\mu e^{\Delta t/\tau_-} & \text{if } \Delta t < 0 \end{cases}$$

for $\Delta t = t_i^{spike} - t_j^{spike} = t_{post} - t_{pre}$ which is positive if $j$ fires before $i$, $A_+ = 0.035$, $A_- = 0.035$ (unless otherwise stated), $\tau_+ = 20$ ms, $\tau_- = 20$ ms (unless otherwise stated), and $\mu = 0.1$. The parameter $\mu$ modulates the update rule between additive ($\mu = 0$) and multiplicative STDP ($\mu = 1$) (Gütig et al. 2003). Additive STDP has the advantage of allowing the weights to explore more of the allowed range of values (Babadi and Abbott 2013). However, it has a couple drawbacks. First, it can generate bi-modal weight distributions of extreme values which are sensitive to changes in the firing rates of pre- and post-synaptic neurons (Rubin et al. 2001). Second, it requires the use of a hard boundary condition ($w_{ij} \rightarrow w_{max}$ if $w_{ij} > w_{max}$). The soft boundary conditions of multiplicative STDP does not suffer from these disadvantages but it limits the dynamics of the weights. Here we use $\mu = 0.1$ which blends the advantages of the two (Gilson and Fukai 2011).

We assume that the change in weight $w_{ij}$ from pre-synaptic neuron $j$ to post-synaptic neuron $i$ sums linearly if $j$ fires multiple times shortly before $i$ fires. Thus, in the simulation, the cSTDP learning rule is implemented algorithmically as follows.

$$\text{cSTDP: } \Delta w_{ij}(t) = \begin{cases} \left(1-w_{ij}(t-1)\right)^\mu P(j,t) & \text{if } i \text{ fires} \\ w_{ij}(t-1)^\mu M(i,t) & \text{if } j \text{ fires} \end{cases}$$

$$\text{rSTDP: } \Delta w_{ij}(t) = \begin{cases} -w_{ij}(t-1)^\mu P(j,t) & \text{if } i \text{ fires} \\ -\left(1-w_{ij}(t-1)\right)^\mu M(i,t) & \text{if } j \text{ fires} \end{cases}$$

where $P(j,t)$ and $M(i,t)$ are an exponentially decaying functions with time constants $\tau_+$ and $\tau_-$ respectively. $P(j,t)$ is increased by $A_+$ when $j$ fires, and $M(i,t)$ is decreased by $A_-$ when $i$ fires. $P(j,t)$ and $M(i,t)$, being functions of neurons, not connections, are independent of STDP type.

These equations implement the STDP exponentials. To illustrate this, consider the case when pre-synaptic neuron $j$ fires (possibly multiple times) before post-synaptic neuron $i$. Each time $j$ fires, $P(j,t)$ is increased by A+ and decays exponentially. Thus, at time t, P (j, t) is the sum of exponential residues of the STDP potentiation curve due to all the spikes pre- synaptic neuron $j$ fired before time $t$. In other words, $P(j,t)$ is the convolution of the positive half of the STDP curve with pre-synaptic neuron $j$'s spike train up until time $t$. Therefore, when post-synaptic neuron $i$ fires at time $t$, the weight $w_{ij}$ updated by $P(j,t)$ reflects the sun total change of STDP due to the interaction of $i$'s action potential with all of the pre-synaptic neuron $j$'s prior action potentials.

We model the weights within and between layers as obeying either cSTDP or rSTDP. In most simulations, all the projections between two layers follow the same learning rule. For example, all the connections from layer 4 to layer 2/3 follow cSTDP or all of those connections follow rSTDP. There are 9 different types of connections: L4 → L4, L2/3 → L4, L5/6 → L4, L4 → L2/3, L2/3 → L2/3, L5/6 → L2/3, L4 → L5/6, L2/3 → L5/6, L5/6 → L5/6. This gives a total of $2^9 = 512$ possible STDP configurations. In Fig. 4a-b and Supplementary Fig. S4, we examine scenarios where $x\%$ of the connections between two layers follow one rule and $(100-x)\%$ follow the other rule. Simulations were run for 60s of simulation time to allow the matrix of average weights to converge (Supplementary Fig. S2). We ran each simulation 5 times with identical parameters except for the noisy input through external Poisson neurons.

**Statistics and analysis**

While weights changed dynamically throughout the simulations, they largely hovered around mean values towards the end of the simulations. Examples of the dynamic changes in individual weights throughout the whole simulation are provided in Supplementary Fig. S2a. Additionally, Supplementary Fig. S2b,c shows the dynamic changes in the weights averaged across all pairs of neurons within each specific pair of layers. To evaluate the degree of convergence in the simulations, we computed the final weight variation defined as the standard deviation of individual weights over the last 5 seconds of the simulation. Histograms of final weight variation for the example STDP configuration used in Fig. 2 are shown in Supplementary Fig. S2d. Simulations showed that, on average, the final weight variation remained small (Supplementary Fig. S2e).

In the analyses of the results, we averaged each individual weight $w$ over the last 5 seconds of the simulation. We show the distribution of all individual weights for each pair of layers for two example configurations in Fig. 2a4, b4. Next, we compute the average across all neuron pairs to build a weight matrix $W$ that has 9 entries (e.g., Fig. 2a3, b3). Averaging is justified by unimodal weight distributions (e.g. Fig. 2a4). Note $W$ denotes average weight matrices while $w$ denotes individual weights.

To evaluate the output of each model, we compared the resulting weight matrices with an idealized target matrix $T$, defined in Fig. 1c, which is a simplification of a canonical inter-laminar connectivity observed in neocortical circuits of macaques and cats (Callaway 1998a, Douglas and Martin 2004). The average weight matrix for each configuration was scored against the binary target weight matrix $T$ using a scaled version of the Frobenius norm while ignoring the diagonal elements. Model success is defined as

$$s = 1 - \sqrt{\frac{1}{6} \sum_{i \neq j} (T_{ij} - W_{ij})^2}$$

where diagonal elements were ignored as not to make any assumptions about the distributions of weights between neurons within the same layer in the target circuit. Note that model success is bounded between 0 and 1 with $s = 1$ if and only if $= T$. We averaged the model success across simulations and ranked the different STDP models according to success.

The best 16 models as ranked by success shared the same STDP configurations at many of the connections. We therefore focused on these configurations and investigated how the success of the best 16 configurations changed with modifications to key model parameters. All of the parameters used in the simulations are shown in Supplementary Table S1 with their default and varied values for testing robustness. Specifically, we varied the ratio of the amounts of excitatory input into each layer, the ratio between STDP parameters $A_-/A_+$ and $\tau_-/\tau_+$, synaptic transmission delays (default = 0 ms), and the percentage of rSTDP and cSTDP in connections between layers.

**Acknowledgements**

We thank James Kozloski and Guillermo Cecchi for providing code from a previous manuscript as well as Emily Mackevicius, Haim Sompolinsky, and German Parisi for discussions and comments on the manuscript. This work was supported by an NSF grant.

# References

Abbott LF, Nelson SB. 2000. Synaptic plasticity: Taming the beast. Nat Neurosci. 3:1178–1183.

Allendoerfer KL, Shatz CJ. 1994. The subplate, a transient neocortical structure: Its role in the development of connections between thalamus and cortex. Annu Rev Neurosci. 17:185–218.

Babadi B, Abbott LF. 2013. Pairwise analysis can account for network structures arising from spike-timing dependent plasticity. PLoS Comput Biol. 9:e1002906.

Bennett JE, Bair W. 2015. Refinement and pattern formation in neural circuits by the interaction of traveling waves with spike-timing dependent plasticity. PLoS Comput Biol, 11:e1004422.

Binzegger T, Douglas RJ, Martin KA. 2004. A quantitative map of the circuit of cat primary visual cortex. Journal of Neuroscience, 24(39), 8441-8453.

Bi G, Poo M. 1998. Synaptic modifications in cultured hippocampal neurons: Dependence on spike timing, synaptic strength, and postsynaptic cell type. J Neurosci. 18:10464–10472.

Bolz J, Castellani V, Mann F, Henke-Fahle S. 1996. Specification of layer-specific connections in the developing cortex. Prog Brain Res. 108:41–54.

Burbank KS, Kreiman G. 2012. Depression-biased reverse plasticity rule is required for stable learning at top-down connections. PLoS Comput Biol. 8:e1002393.

Butts DA, Kanold PO, Shatz CJ. 2007. A burst-based "hebbian" learning rule at retinogeniculate synapses links retinal waves to activity-dependent refinement. PLoS Biol. 5:e61.

Callaway EM. 1998a. Local circuits in primary visual cortex of the macaque monkey. Annu Rev Neurosci. 21:47–74.

Callaway EM. 1998b. Prenatal development of layer-specific local circuits in primary visual cortex of the macaque monkey. J Neurosci. 18:1505–1527.

Caporale N, Dan Y. 2008. Spike timing–dependent plasticity: A hebbian learning rule. Annu Rev Neurosci. 31:25–46.

Castellani V, Bolz J. 1997. Membrane-associated molecules regulate the formation of layer-specific cortical circuits. Proc Natl Acad Sci. 94:7030–7035.

Constantinople CM, Bruno RM. 2013. Deep cortical layers are activated directly by thalamus. Science. 340:1591–1594.

Debanne D, Gähwiler BH, Thompson SM. 1998. Long-term synaptic plasticity between pairs of individual ca3 pyramidal cells in rat hippocampal slice cultures. J Physiol. 507:237–247.

Douglas RJ, Martin KA. 2004. Neuronal circuits of the neocortex. Annu Rev Neurosci. 27:419–451.

Egger V, Feldmeyer D, Sakmann B. 1999. Coincidence detection and changes of synaptic efficacy in spiny stellate neurons in rat barrel cortex. Nat Neurosci. 2:1098–1105.

Espinosa JS, Stryker MP. 2012. Development and plasticity of the primary visual cortex. Neuron. 75:230–249.

Feldman DE. 2000. Timing-based ltp and ltd at vertical inputs to layer ii/iii pyramidal cells in rat barrel cortex. Neuron. 27:45–56.

Feldman DE, Brecht M. 2005. Map plasticity in somatosensory cortex. Science. 310:810–815.

Felleman DJ, Van Essen DC. 1991. Distributed hierarchical processing in the primate cerebral cortex. Cereb Cortex. 1:1–47.

Fox K, Wong RO. 2005. A comparison of experience-dependent plasticity in the visual and somatosensory systems. Neuron. 48:465–477.

Froemke RC, Letzkus JJ, Kampa BM, Hang GB, Stuart GJ. 2010. Dendritic synapse location and neocortical spike-timing-dependent plasticity. Front Synaptic Neurosci. 2.

Froemke RC, Poo M, Dan Y. 2005. Spike-timing-dependent synaptic plasticity depends on dendritic location. Nature. 434:221–225.

Gilson M, Fukai T. 2011. Stability versus neuronal specialization for STDP: long-tail weight distributions solve the dilemma. PloS One. 6:e25339.

Gütig R. Aharonov R, Rotter S, Sompolinsky H. 2003. Learning input correlations through nonlinear temporally asymmetric Hebbian plasticity. The Journal of Neuroscience. 23(9), 3697-3714.

Karmarkar UR, Dan Y. 2006. Experience-dependent plasticity in adult visual cortex. Neuron. 52:577–585.

Kozloski J, Cecchi GA. 2010. A theory of loop formation and elimination by spike timing-dependent plasticity. Front Neural Circuits. 4.

Larsen DD, Callaway EM. 2006. Development of layer-specific axonal arborizations in mouse primary somatosensory cortex. J Comp Neurol. 494:398–414.

Letzkus JJ, Kampa BM, Stuart GJ. 2006. Learning rules for spike timing-dependent plasticity depend on dendritic

synapse location. J Neurosci. 26:10420–10429.

Li H, Fertuzinhos S, Mohns E, Hnasko, TS, Verhage M, Edwards R, Sestan N, Crair MC. 2013. Laminar and columnar development of barrel cortex relies on thalamocortical neurotransmission. Neuron. 79:970–986.

Lim S, McKee JL, Woloszyn L, Amit Y, Freedman DJ, Sheinberg DL, Brunel N. 2015. Inferring learning rules from distributions of firing rates in cortical neurons. Nat Neurosci. 18:1804–1810.

Lui JH, Hansen DV, Kriegstein AR. 2011. Development and evolution of the human neocortex. Cell. 146:18–36.

Lund R, Mustari M. 1977. Development of the geniculocortical pathway in rat. J Comp Neurol. 173:289–305.

Markram H, Lübke J, Frotscher M, Sakmann B. 1997. Regulation of synaptic efficacy by coincidence of postsynaptic aps and epsps. Science. 275:213–215.

Miller KD. 2003. Understanding layer 4 of the cortical circuit: A model based on cat v1. Cereb Cortex. 13:73–82.

Miyashita-Lin EM, Hevner R, Wassarman KM, Martinez S, Rubenstein JL. 1999. Early neocortical regionalization in the absence of thalamic innervation. Science. 285:906–909.

Rakic P. 1977. Prenatal development of the visual system in rhesus monkey. Philos Trans R Soc Lond B Biol Sci. 278:245–260.

Rakic P. 2009. Evolution of the neocortex: A perspective from developmental biology. Nat Rev Neurosci. 10:724–735.

Rubin J, Lee DD, Sompolinsky H. 2001. Equilibrium properties of temporally asymmetric hebbian plasticity. Phys Rev Lett. 86:364.

Shatz CJ, Luskin MB. 1986. The relationship between the geniculocortical afferents and their cortical target cells during development of the cat's primary visual cortex. J Neurosci. 6:3655–3668.

Silbereis JC, Pochareddy S, Zhu Y, Li M, Sestan N. 2016. The cellular and molecular landscapes of the developing human central nervous system. Neuron. 89:248–268.

Sjöström PJ, Häusser M. 2006. A cooperative switch determines the sign of synaptic plasticity in distal dendrites of neocortical pyramidal neurons. Neuron. 51:227–238.

Sjöström PJ, Turrigiano GG, Nelson SB. 2001. Rate, timing, and cooperativity jointly determine cortical synaptic plasticity. Neuron. 32:1149–1164.

Song S, Miller KD, Abbott LF. 2000. Competitive hebbian learning through spike-timing-dependent synaptic plasticity. Nat Neurosci. 3:919–926.

Wess JM, Isaiah A, Watkins PV, Kanold PO. 2017. Subplate neurons are the first cortical neurons to respond to sensory stimuli. Proc Natl Acad Sci. 114:12602–12607.

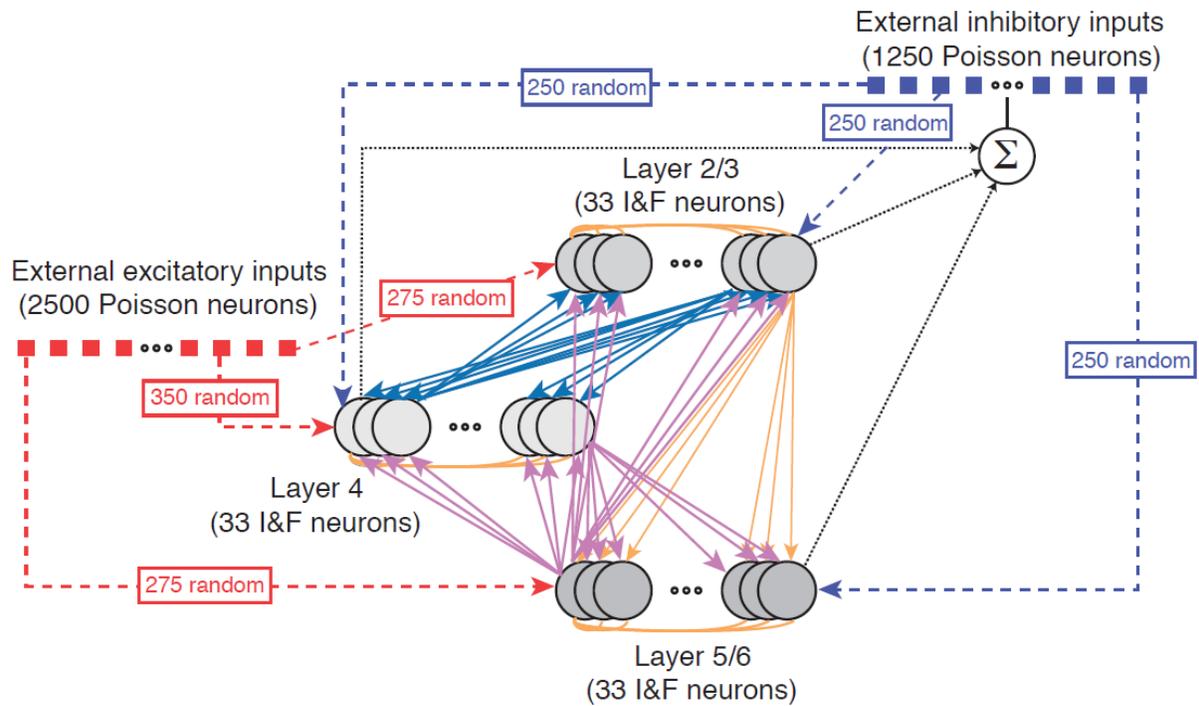

Figure S1
**Schematic illustration of model architecture.** The model consists of 3 layers, each one with 33 neurons, plus external excitatory inputs (red squares) and external inhibitory inputs (blue squares). Neurons are initially connected in an all-to-all fashion, only some of the representative connections are rendered here for pictorial clarity. The color of the connections corresponds to the colors and proposed learning rules in Fig. 3.

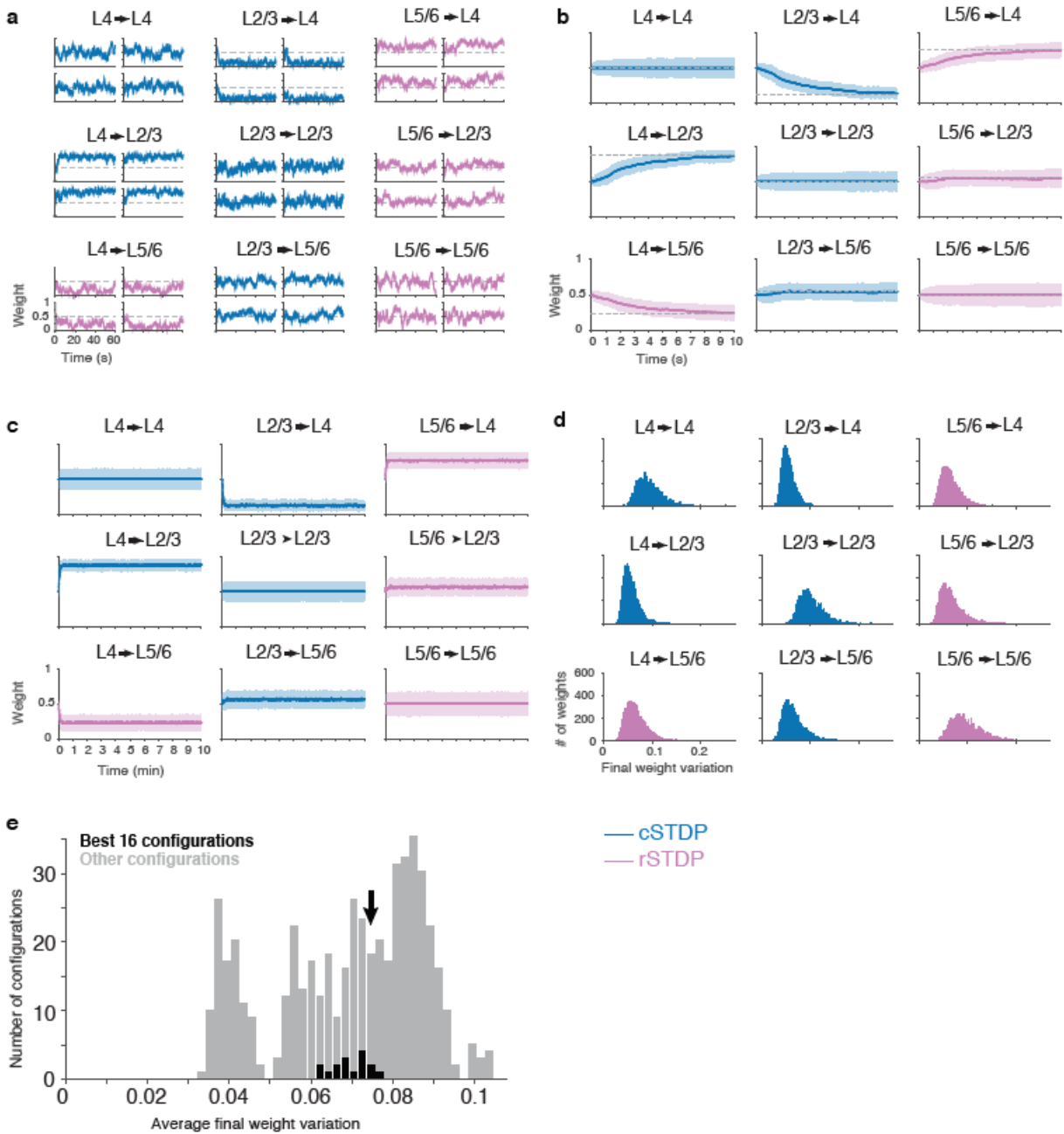

Figure S2

**Convergence of simulations. a,** Example dynamics of individual weights from the configuration in Fig. 2a. For each pair of layers, the plots follow 4 random example weights over the 60 seconds of simulation. The dashed lines indicate the initial conditions. **b,** Dynamics during the first 10 seconds, showing the average of all weights for each pair of layers from a single simulation and for the same configuration as in (**a**). The shaded areas denote 1 SD and $n = 5{,}445$. **c,** Dynamics during 10 minutes, showing the average of all weights for each pair of layers from a single simulation and for the same configuration as in (**a**). The shaded areas denote 1 SD and $n = 5{,}445$. **d,** Histograms showing distribution of final weight variation (standard deviation of the weights over the last 5 seconds of the simulation) for the same configuration in (**a**) and across 5 simulations ($n = 5{,}445$). **e,** Average of final weight variation for each of the 512 configurations. The best 16 configurations are highlighted in black and the example from (**a-c**) is labeled by an arrow.

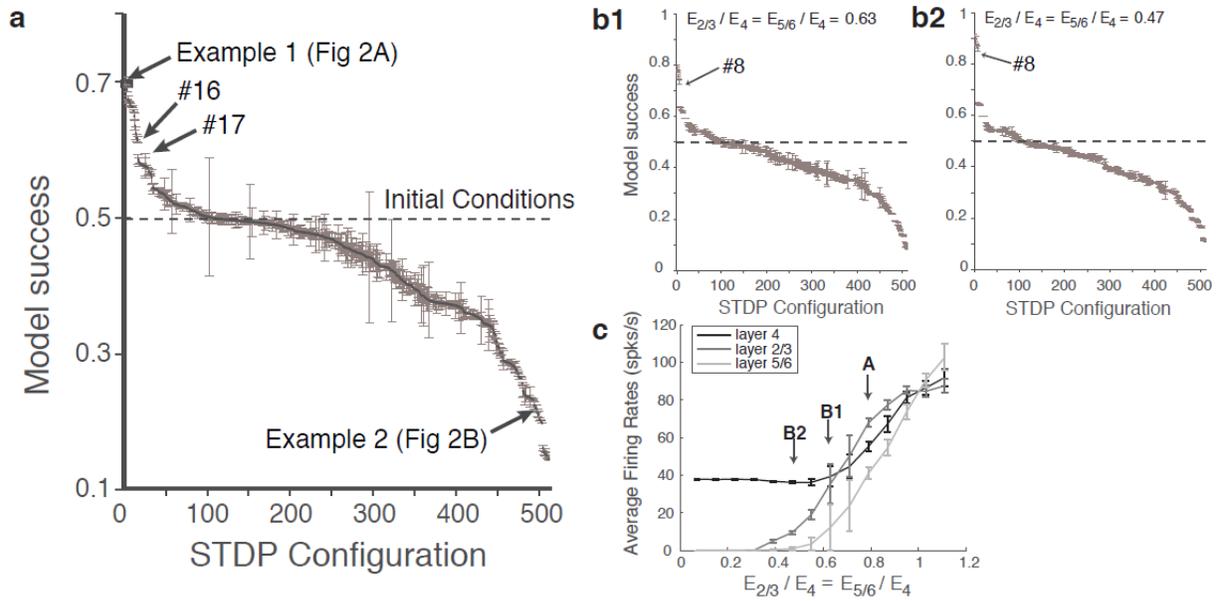

Figure S3
**Model success for all possible configurations. a,** The y-axis shows the model success (defined in the text), 0.5 is the success of the initial conditions (horizontal dashed line). Model success is averaged over 5 simulations. Error bars denote 1 SD. Example 1 is the configuration shown in Fig. 2a and Example 2 is the configuration shown in Fig. 2b. Note the gap between configuration number 16 and configuration number 17, as well as the gap before the bottom 8 simulations. **b,** Model success (mean $\pm$ 1 SD with $n = 5$), for all possible configurations with $E_{2/3}/E_4 = E_{5/6}/E_4 = 0.63$ (left), 0.47 (right). In (**a**), $E_{2/3}/E_4 = E_{5/6}/E_4 = 0.79$. Note that as the excitatory input ratio decreases, a large gap emerges between configuration 8 and 9 and the gap between 16 and 17 grows. **c,** The average firing of the top 16 (averaged over 5 simulations as well as the 16 configurations) separated by layer as a function of $E_{2/3}/E_4 = E_{5/6}/E_4$. Firing rates are averaged over the last 10 seconds of simulation time. Note that although $E_{2/3}/E_4 = E_{5/6}/E_4 = 0.47$ shows a more defined "top 16", the simulations result in a network with unrealistically low firing rates in layer 5/6.

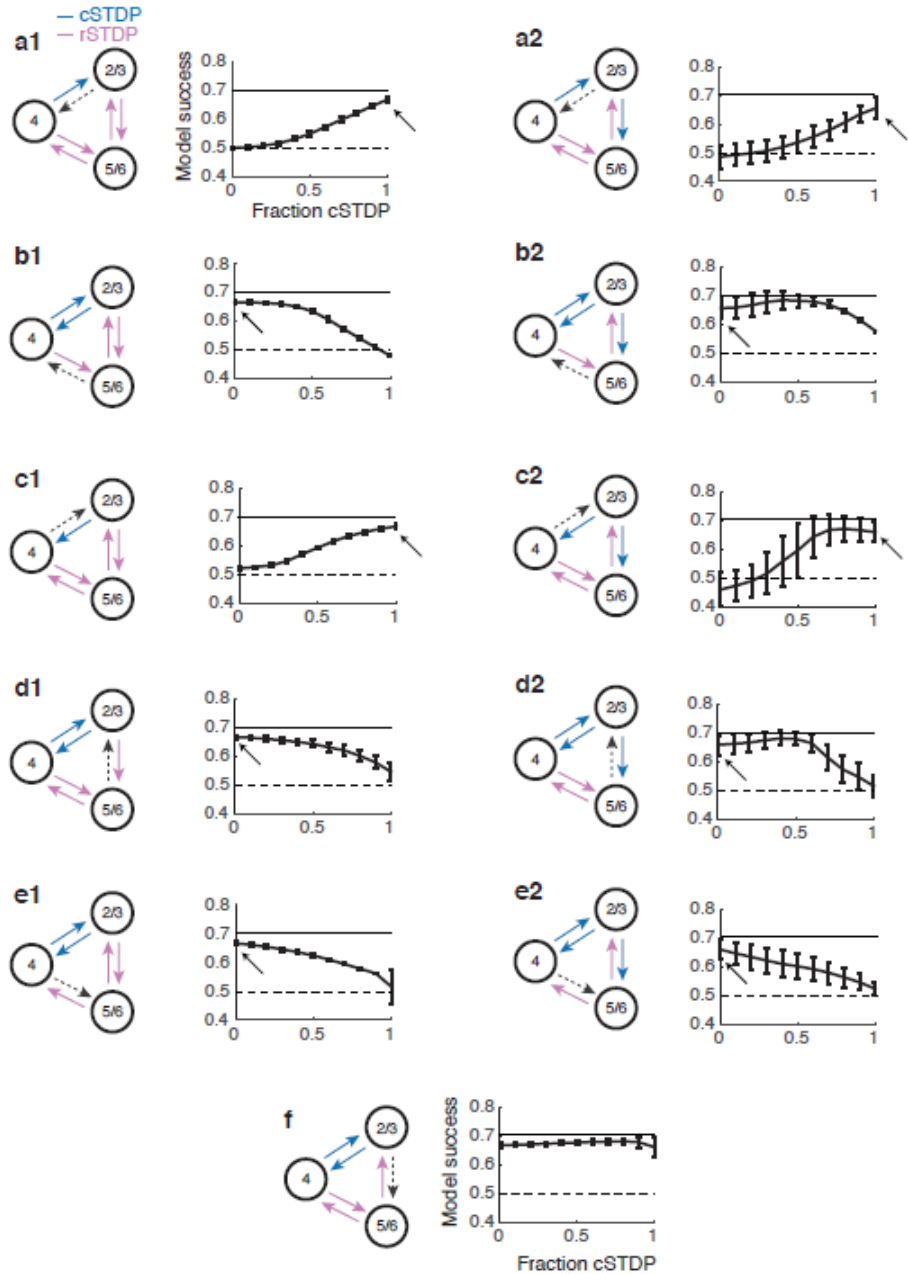

Figure S4
**Performance of hybrid models combining cSTDP and rSTDP.** Following the procedure illustrated in Fig. 4a-b, one of the connections is allowed to have a mixture of cSTDP and rSTDP (dashed arrow in model scheme) while all the other connections keep the configuration in Fig. 3 (fraction of cSTDP = 0 indicates all weights follow rSTDP and fraction = 1 indicates that all weights follow cSTDP). The y axis shows the model success, averaged over 5 simulations and across within-layer connections (8 possible configurations) for a total of $n = 40$; error bars denote 1 SD. The horizontal dashed line shows the initial conditions (success = 0.5) and the dotted lines shows the success of the best configuration. The arrow indicates the configuration in Fig. 3. The left column shows models where the connection from L2/3 to L5/6 has rSTDP and the right column shows models with cSTDP for that connection. Part (**d1**) is identical to Fig. 4a and part (**f**) is identical to Fig. 4b, and they are reproduced here for completeness.

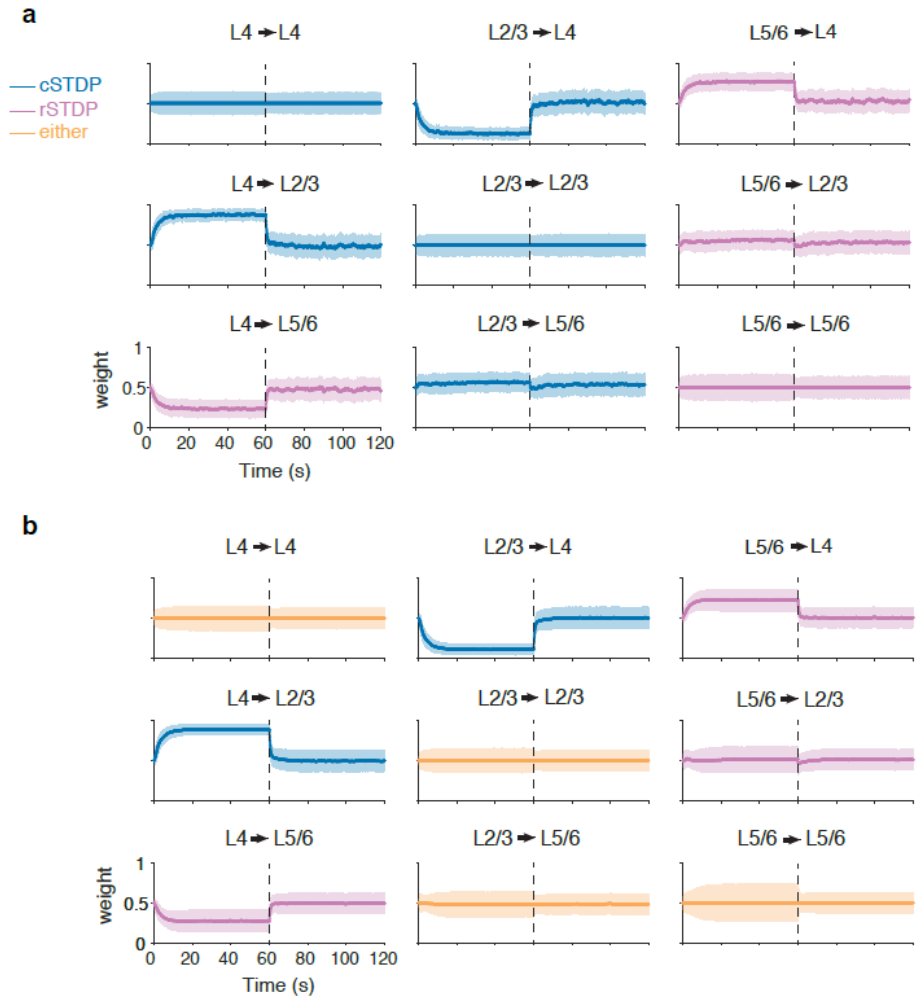

Figure S5
**Weight dynamics when external input is switched to being equal for all layers.** After 60 seconds of simulation (dashed line), the amount of external input is changed from the default values to $E_4 = E_{2/3} = E_{5/6} = 350$. Shown are the average of all weights for each pair of layers. . Error bars denote 1 SD. **a,** Example dynamics of weights from the configuration in Fig. 2a ($n = 5,445$). **b,** Example dynamics of weights for the best 16 configurations ($n = 87,120$).

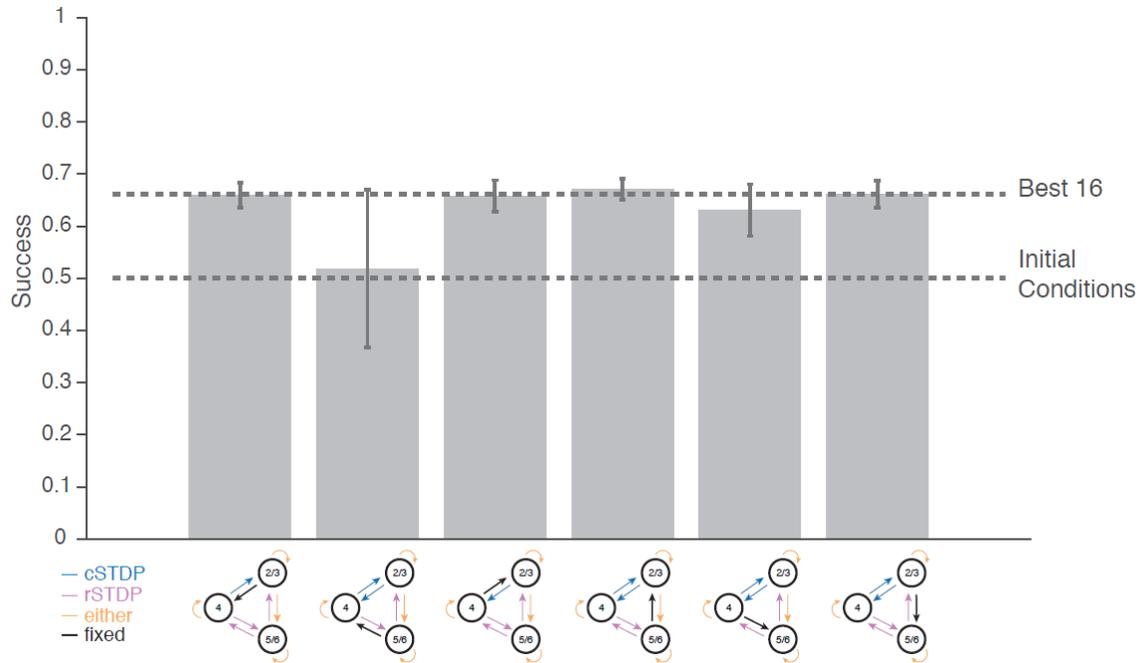

Figure S6
**Success of best 16 configurations when one inter-laminar connection develops first.** In these simulations, one of the inter-laminar connections (shown in black) is fixed from the beginning to the weight values corresponding to the value reported in Fig. 3c (final averages for the 16 best configurations). All the remaining connections are initialized and undergo STDP as in the default simulations. Error bars denote 1 SD ($n = 80$).

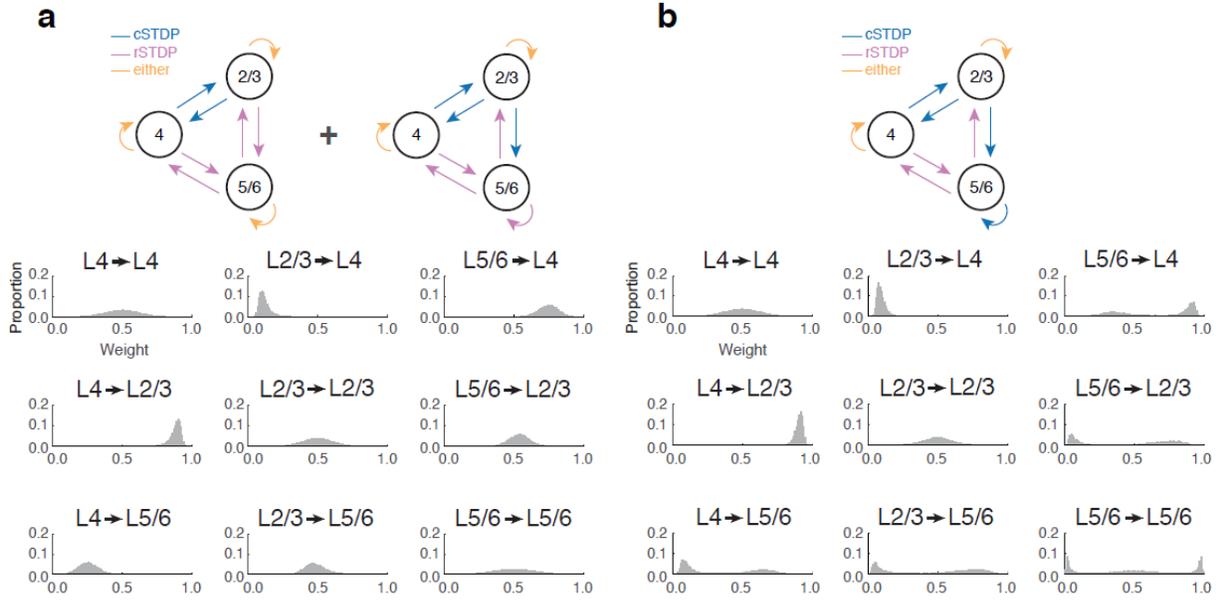

Figure S7
**Multimodality within the best 16 configurations. a,** Weight histograms from the 12 best configurations that display unimodal weight distributions, pooled across 5 simulations for a total $n = 60$. **b,** Weight histograms from the 4 best configurations that display multi-modal weight distributions in connections into and out of layer 5/6 ($n = 20$).

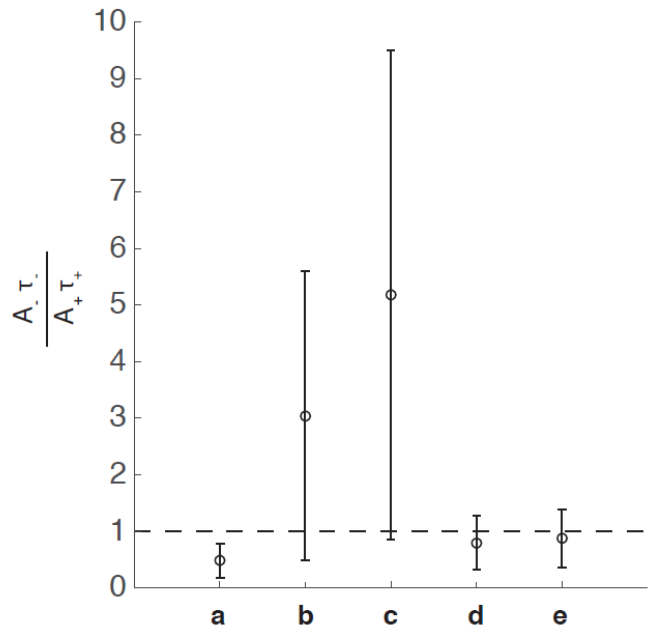

Figure S8
**Experimentally estimated balance between potentiation and depression. Experimentally estimated balance between potentiation and depression. a,** Data from proximal synapses along the apical dendrite of layer 2/3 pyramidal neurons in rat visual cortex (Froemke et al. 2005). **b,** Data from distal synapses along the apical dendrite of layer 2/3 (Froemke et al. 2005). **c,** Data from vertical inputs to layer 2/3 pyramidal neurons of rat S1 (Feldman 2000). **d,** Data from glutamatergic synapses from dissociated rat hippocampal neurons (Bi and Poo 1998). **e,** Data of retinal neuron's synapses onto optic tectum neurons in *Xenopus* tadpoles (Zhang et al. 1998). Note about the calculations. We extracted the change in synaptic strength values as a function of time between spikes from the corresponding figures. We estimated $A$ and $\tau$ by fitting the curves with an exponential function (Matlab's "fit" function) and used the fitted values. The figure shows the average values and the error bars were calculated by error propagation.

**References:**
Bi GQ, Poo MM. 1998. Synaptic modifications in cultured hippocampal neurons: dependence on spike timing, synaptic strength, and postsynaptic cell type. J Neurosci. 18:10464-10472.
Feldman DE. 2000. Timing-based LTP and LTD at vertical inputs to layer II/III pyramidal cells in rat barrel cortex. Neuron. 27:45-56.
Froemke RC, Poo MM, Dan Y. 2005. Spike-timing-dependent synaptic plasticity depends on dendritic location. Nature. 434:221.
Zhang LI, Tao HW, Holt CE, Harris WA, Poo MM. 1998. A critical window for cooperation and competition among developing retinotectal synapses. Nature. 395:37.

| Simulation Parameters | Description | Default value | Values explored |
|---|---|---|---|
| $dt$ | Simulation time step | 0.1 ms | |
| $T$ | Length of simulation | 60 sec | |
| | | | |
| **STDP parameters** | **Description** | **Default value** | **Values explored** |
| $A_-$ | STDP amplitude for $\Delta t < 0$ | $3.5 * 10^{-2}$ | $[0.7, 7]*10^{-2}$ |
| $A_+$ | STDP amplitude for $\Delta t > 0$ | $3.5 * 10^{-2}$ | |
| $\tau_-$ | Time-constant for $\Delta t < 0$ | 20 ms | $[4, 40]$ ms |
| $\tau_+$ | Time-constant for $\Delta t > 0$ | 20 ms | |
| $\mu$ | Governs additive vs multiplicative STDP | 0.1 | |
| $w_{max}$ | Maximum weight value for inter-network | 1 | |
| $w_{min}$ | Minimum weight value for inter-network | 0 | |
| $w_{inh}$ | Fixed weight of inhibitory connections | 1.5 | |
| $STDP_{mod}$ | Percent of connections for a given layer that are cSTDP | Either 0 or 1 | $[0, 1]$ |
| | | | |
| **Neuron Parameters** | **Description** | **Default value** | **Values explored** |
| $\tau_m$ | Decay time-constant for membrane potential | 20 ms | |
| $\tau_{exc}$ | Decay time-constant for voltage potential at excitatory synapses | 5 ms | |
| $\tau_{inh}$ | Decay time-constant for voltage potential at inhibitory synapses | 5 ms | |
| $V_{rest}$ | Resting membrane potential | -60 mV | |
| $V_{thresh}$ | Threshold membrane potential | -54 mV | |
| $V_{reset}$ | Membrane potential reset value following a action potential | -60 mV | |
| $E_{exc}$ | Excitatory reversal potential | 0 mV | |
| $E_{inh}$ | Inhibitory reversal potential | -70 mV | |
| | | | |
| **Network Parameters** | **Description** | **Default value** | **Values explored** |
| $N_{neurons}$ | Number of neurons in the network | 99 | |
| $N_{Exc}$ | Number of extra-network excitatory homogenous Poisson neurons | 2500 | |
| $E_4$ | Number of extra-network excitatory Poisson neurons projecting to layer 4 | 350 | |
| $E_{2/3}$ | Number of extra-network excitatory Poisson neurons projecting to layer 2/3 | 275 | $[0, 389]$ |
| $E_{5/6}$ | Number of extra-network excitatory Poisson neurons projecting to layer 5/6 | 275 | $[0, 389]$ |
| $r_{exc}$ | Firing rate for excitatory Poisson neurons | 20 Hz | |
| $N_{Inh}$ | Number of fast inhibition non-homogenous Poisson neurons | 1250 | |
| $I_{layer}$ | Number of fast inhibition Poisson neurons projecting to each layer | 250 | |
| $\tau_I$ | Firing rate time constant for inhibitory Poisson neurons | 2 ms | |
| $r_{inh}^{max}$ | Maximum firing rate for inhibitory Poisson neurons | 1000 Hz | |
| $r_{inh}^{min}$ | Minimum firing rate for inhibitory Poisson neurons | 5 Hz | |
| $D_{intra}$ | Synaptic delay between neurons within the same layer | 0 ms | $[0, 5]$ ms |
| $D_{inter}$ | Synaptic delay between neurons in different layers | 0 ms | $[0, 5]$ ms |

Table S1
**Parameters used in the simulations.** This table lists all the parameters used in the simulations, the corresponding default values and the range of values explored for some of them when evaluating robustness to parameter changes (see text for further details). The interval step used for varying parameters are: $0.175 \times 10^{-2}$ for $A_-$; 1 ms for $\tau_-$; 0.1 for $STDP_{mod}$; 28 for $E_{2/3}$ (for values 25 through 389); 28 for $E_{5/6}$ (for values 25 through 389); 1 ms for $D_{intra}$ and $D_{inter}$.

| T   | L2/3→L4     | L5/6→L4     | L4→L2/3     | L5/6→L2/3   | L4→L5/6     | L2/3→L5/6   | Success     |
|-----|-------------|-------------|-------------|-------------|-------------|-------------|-------------|
|     | 0.00        | 1.00        | 1.00        | 1.00        | 0.00        | 1.00        |             |
| 1   | 0.12 ± 0.03 | 0.76 ± 0.08 | 0.88 ± 0.03 | 0.56 ± 0.09 | 0.24 ± 0.08 | 0.56 ± 0.09 | 0.70 ± 0.01 |
| 2   | 0.15 ± 0.05 | 0.74 ± 0.08 | 0.85 ± 0.05 | 0.56 ± 0.09 | 0.26 ± 0.08 | 0.56 ± 0.09 | 0.69 ± 0.00 |
| 3   | 0.10 ± 0.02 | 0.77 ± 0.08 | 0.90 ± 0.02 | 0.53 ± 0.10 | 0.23 ± 0.08 | 0.53 ± 0.10 | 0.69 ± 0.00 |
| 4   | 0.10 ± 0.03 | 0.77 ± 0.05 | 0.90 ± 0.03 | 0.53 ± 0.06 | 0.23 ± 0.05 | 0.47 ± 0.06 | 0.68 ± 0.00 |
| 5   | 0.12 ± 0.04 | 0.74 ± 0.09 | 0.88 ± 0.04 | 0.52 ± 0.10 | 0.26 ± 0.09 | 0.52 ± 0.10 | 0.68 ± 0.01 |
| 6   | 0.10 ± 0.03 | 0.75 ± 0.05 | 0.90 ± 0.03 | 0.52 ± 0.07 | 0.25 ± 0.05 | 0.48 ± 0.07 | 0.67 ± 0.00 |
| 7   | 0.11 ± 0.03 | 0.75 ± 0.06 | 0.89 ± 0.03 | 0.56 ± 0.05 | 0.25 ± 0.06 | 0.44 ± 0.05 | 0.67 ± 0.00 |
| 8   | 0.12 ± 0.04 | 0.74 ± 0.06 | 0.88 ± 0.04 | 0.55 ± 0.06 | 0.26 ± 0.06 | 0.45 ± 0.06 | 0.67 ± 0.00 |
| 9   | 0.13 ± 0.04 | 0.74 ± 0.06 | 0.87 ± 0.04 | 0.53 ± 0.06 | 0.26 ± 0.06 | 0.47 ± 0.06 | 0.67 ± 0.00 |
| 10  | 0.12 ± 0.04 | 0.73 ± 0.06 | 0.88 ± 0.04 | 0.51 ± 0.07 | 0.27 ± 0.06 | 0.49 ± 0.07 | 0.67 ± 0.00 |
| 11  | 0.13 ± 0.05 | 0.72 ± 0.06 | 0.87 ± 0.05 | 0.55 ± 0.05 | 0.28 ± 0.06 | 0.45 ± 0.05 | 0.66 ± 0.00 |
| 12  | 0.15 ± 0.07 | 0.70 ± 0.06 | 0.85 ± 0.07 | 0.55 ± 0.07 | 0.30 ± 0.06 | 0.45 ± 0.07 | 0.65 ± 0.01 |
| 13  | 0.09 ± 0.03 | 0.68 ± 0.27 | 0.91 ± 0.03 | 0.47 ± 0.34 | 0.32 ± 0.27 | 0.47 ± 0.34 | 0.64 ± 0.01 |
| 14  | 0.10 ± 0.03 | 0.66 ± 0.27 | 0.90 ± 0.03 | 0.46 ± 0.34 | 0.34 ± 0.27 | 0.46 ± 0.34 | 0.63 ± 0.01 |
| 15  | 0.09 ± 0.03 | 0.67 ± 0.26 | 0.91 ± 0.03 | 0.44 ± 0.31 | 0.33 ± 0.26 | 0.44 ± 0.31 | 0.62 ± 0.01 |
| 16  | 0.08 ± 0.02 | 0.67 ± 0.28 | 0.92 ± 0.02 | 0.43 ± 0.34 | 0.33 ± 0.28 | 0.43 ± 0.34 | 0.62 ± 0.00 |
| ... |             |             |             |             |             |             |             |
| 241 | 0.57 ± 0.08 | 0.42 ± 0.08 | 0.43 ± 0.08 | 0.44 ± 0.13 | 0.42 ± 0.08 | 0.56 ± 0.13 | 0.47 ± 0.00 |
| 242 | 0.49 ± 0.18 | 0.51 ± 0.28 | 0.49 ± 0.18 | 0.43 ± 0.28 | 0.51 ± 0.28 | 0.43 ± 0.28 | 0.47 ± 0.03 |
| 243 | 0.48 ± 0.11 | 0.53 ± 0.07 | 0.48 ± 0.11 | 0.40 ± 0.18 | 0.47 ± 0.07 | 0.40 ± 0.18 | 0.47 ± 0.02 |
| 244 | 0.42 ± 0.07 | 0.55 ± 0.06 | 0.42 ± 0.07 | 0.38 ± 0.15 | 0.45 ± 0.06 | 0.38 ± 0.15 | 0.47 ± 0.01 |
| 245 | 0.56 ± 0.06 | 0.44 ± 0.06 | 0.44 ± 0.06 | 0.47 ± 0.16 | 0.44 ± 0.06 | 0.47 ± 0.16 | 0.47 ± 0.01 |
| 246 | 0.46 ± 0.07 | 0.57 ± 0.13 | 0.46 ± 0.07 | 0.42 ± 0.12 | 0.57 ± 0.13 | 0.42 ± 0.12 | 0.47 ± 0.01 |
| 247 | 0.54 ± 0.06 | 0.51 ± 0.05 | 0.46 ± 0.06 | 0.44 ± 0.13 | 0.49 ± 0.05 | 0.44 ± 0.13 | 0.47 ± 0.01 |
| 248 | 0.50 ± 0.09 | 0.52 ± 0.05 | 0.50 ± 0.09 | 0.40 ± 0.17 | 0.48 ± 0.05 | 0.40 ± 0.17 | 0.47 ± 0.01 |
| 249 | 0.58 ± 0.07 | 0.42 ± 0.07 | 0.42 ± 0.07 | 0.50 ± 0.10 | 0.42 ± 0.07 | 0.50 ± 0.10 | 0.47 ± 0.00 |
| 250 | 0.58 ± 0.06 | 0.42 ± 0.07 | 0.42 ± 0.06 | 0.56 ± 0.12 | 0.42 ± 0.07 | 0.44 ± 0.12 | 0.47 ± 0.00 |
| 251 | 0.46 ± 0.08 | 0.52 ± 0.11 | 0.46 ± 0.08 | 0.40 ± 0.18 | 0.48 ± 0.11 | 0.40 ± 0.18 | 0.47 ± 0.01 |
| 252 | 0.57 ± 0.13 | 0.47 ± 0.07 | 0.57 ± 0.13 | 0.42 ± 0.13 | 0.47 ± 0.07 | 0.42 ± 0.13 | 0.47 ± 0.00 |
| 253 | 0.57 ± 0.05 | 0.52 ± 0.12 | 0.43 ± 0.05 | 0.64 ± 0.20 | 0.52 ± 0.12 | 0.36 ± 0.20 | 0.47 ± 0.00 |
| 254 | 0.54 ± 0.05 | 0.46 ± 0.11 | 0.46 ± 0.05 | 0.62 ± 0.19 | 0.54 ± 0.11 | 0.38 ± 0.19 | 0.47 ± 0.01 |
| 255 | 0.59 ± 0.15 | 0.46 ± 0.06 | 0.59 ± 0.15 | 0.41 ± 0.12 | 0.46 ± 0.06 | 0.41 ± 0.12 | 0.47 ± 0.01 |
| 256 | 0.54 ± 0.06 | 0.35 ± 0.25 | 0.46 ± 0.06 | 0.70 ± 0.23 | 0.35 ± 0.25 | 0.30 ± 0.23 | 0.47 ± 0.01 |
| ... |             |             |             |             |             |             |             |
| 497 | 0.69 ± 0.06 | 0.09 ± 0.06 | 0.31 ± 0.06 | 0.27 ± 0.09 | 0.91 ± 0.06 | 0.27 ± 0.09 | 0.22 ± 0.01 |
| 498 | 0.01 ± 0.01 | 0.06 ± 0.08 | 0.01 ± 0.01 | 0.99 ± 0.01 | 0.94 ± 0.08 | 0.01 ± 0.01 | 0.21 ± 0.01 |
| 499 | 0.71 ± 0.06 | 0.10 ± 0.08 | 0.29 ± 0.06 | 0.26 ± 0.08 | 0.90 ± 0.08 | 0.26 ± 0.08 | 0.21 ± 0.01 |
| 500 | 0.01 ± 0.00 | 0.05 ± 0.09 | 0.01 ± 0.00 | 0.99 ± 0.01 | 0.95 ± 0.09 | 0.01 ± 0.01 | 0.21 ± 0.01 |
| 501 | 0.01 ± 0.00 | 0.04 ± 0.02 | 0.01 ± 0.00 | 0.99 ± 0.00 | 0.96 ± 0.02 | 0.01 ± 0.00 | 0.20 ± 0.00 |
| 502 | 0.01 ± 0.00 | 0.03 ± 0.02 | 0.01 ± 0.00 | 0.98 ± 0.01 | 0.97 ± 0.02 | 0.02 ± 0.01 | 0.20 ± 0.00 |
| 503 | 0.01 ± 0.00 | 0.03 ± 0.01 | 0.01 ± 0.00 | 0.99 ± 0.00 | 0.97 ± 0.01 | 0.01 ± 0.00 | 0.20 ± 0.00 |
| 504 | 0.01 ± 0.00 | 0.02 ± 0.01 | 0.01 ± 0.00 | 0.99 ± 0.01 | 0.98 ± 0.01 | 0.01 ± 0.01 | 0.20 ± 0.00 |
| 505 | 0.07 ± 0.04 | 0.09 ± 0.11 | 0.07 ± 0.04 | 0.08 ± 0.05 | 0.91 ± 0.11 | 0.08 ± 0.05 | 0.16 ± 0.00 |
| 506 | 0.05 ± 0.03 | 0.05 ± 0.06 | 0.05 ± 0.03 | 0.11 ± 0.06 | 0.95 ± 0.06 | 0.11 ± 0.06 | 0.15 ± 0.00 |
| 507 | 0.07 ± 0.04 | 0.08 ± 0.08 | 0.07 ± 0.04 | 0.08 ± 0.04 | 0.92 ± 0.08 | 0.08 ± 0.04 | 0.15 ± 0.00 |
| 508 | 0.05 ± 0.03 | 0.05 ± 0.05 | 0.05 ± 0.03 | 0.10 ± 0.06 | 0.95 ± 0.05 | 0.10 ± 0.06 | 0.15 ± 0.00 |
| 509 | 0.04 ± 0.02 | 0.04 ± 0.02 | 0.04 ± 0.02 | 0.12 ± 0.06 | 0.96 ± 0.02 | 0.12 ± 0.06 | 0.15 ± 0.00 |
| 510 | 0.04 ± 0.02 | 0.04 ± 0.02 | 0.04 ± 0.02 | 0.11 ± 0.07 | 0.96 ± 0.02 | 0.11 ± 0.07 | 0.15 ± 0.00 |
| 511 | 0.06 ± 0.03 | 0.04 ± 0.02 | 0.06 ± 0.03 | 0.08 ± 0.04 | 0.96 ± 0.02 | 0.08 ± 0.04 | 0.14 ± 0.00 |
| 512 | 0.06 ± 0.03 | 0.05 ± 0.02 | 0.06 ± 0.03 | 0.08 ± 0.04 | 0.95 ± 0.02 | 0.08 ± 0.04 | 0.14 ± 0.00 |

Table S2

**Weights and success metric for best, middle, and worst 16 configurations.** The first column indicates the configuration number from 1 through 512, ranked based on the success metric (the first row labeled $T$ depicts the target values). Each successive column indicates one of the 6 between layer connection types (described at the top). Average weights +/- 1 SD are shown for each configuration and connection type (averaged across all neurons between the pair of layers and across 5 simulations, $n = 33 \times 33 \times 5 = 5{,}445$). Colors correspond to cSTDP (blue) or rSTDP (pink). Note the high degree of consistency in the learning rules for 5 of the 6 between-layer connections for the best 16 and worst 16 configurations.